**The Social Impact of Generative LLM-Based AI**[*]

Yu Xie

Princeton University and Peking University

Sofia Avila

Princeton University

(Version: October 2024)

---

* Please direct all correspondence to Yu Xie (yuxie@princeton.edu) or Sofia Avila (sofiaavila@princeton.edu). The research was partially supported by the Paul and Marcia Wythes Center on Contemporary China and Office of Population Research at Princeton University. We are grateful to Wen Liu, Gou Wu, and Dean Minello for their excellent research assistance. The ideas expressed herein are those of the authors.

**Abstract**

Liking it or not, ready or not, we are likely to enter a new phase of human history in which Artificial Intelligence (AI) will dominate economic production and social life – the AI Revolution. Before the actual arrival of the AI Revolution, it is time for us to speculate on how AI will impact the social world. In this article, we focus on the social impact of generative LLM-based AI (GELLMAI), discussing societal factors that contribute to its technological development and its potential roles in enhancing both between-country and within-country social inequality. There are good indications that the US and China will lead the field and will be the main competitors for domination of AI in the world. We conjecture the AI Revolution will likely give rise to a post-knowledge society in which knowledge per se will become less important than in today's world. Instead, individual relationships and social identity will become more important. So will soft skills.

## The Social Impact of Generative LLM-Based AI

With the advent of Generative Large Language Model (LLM)-based Artificial Intelligence (AI) tools such as ChatGPT from OpenAI and Bard from Google, it is natural to wonder about the social impact of this technology. In the remainder of this paper, we will refer to generative LLM-based AI simply as GELLMAI. The main objective of this paper is to explore, tentatively, the social impact of GELLMAI.

While the question about the social impact of GELLMAI is undoubtedly important, any answers must be tentative and speculative at this point. We are still in the early stages of GELLMAI and may need to wait years, perhaps even decades, to fully understand its social implications. However, drawing from our experiences with past technologies in history, our current understanding of GELLMAI, empirical knowledge about the social world, and sociological reasoning, we can engage in preliminary and speculative discussions. We offer our account below.

We believe that the social impact of GELLMAI is enormous, with the potential to revolutionize not only the production of goods and services but also to fundamentally alter the organization of human societies and the nature of daily life. Indeed, this technology holds the potential to significantly increase inequality both *between* and *within* countries, and we will discuss each of these in turn. As we explore these topics, we must keep in mind the speculative nature of our analysis, given the limited knowledge currently available about this technology and its capabilities. In writing this article, we draw inspiration from the ambition and style of Daniel Bell's (1973) pioneering book, *The Coming of the Post-Industrial Society*. Bell's work, published long before the actual arrival of the digital age, serves as a model for our discussion of the future implications of GELLMAI as an emerging transformative technology.

Because the potential social impact of GELLMAI is too vast to be comprehensively discussed in a single article and our current knowledge and ideas about this emerging technology are still evolving, we will focus and structure our discussion as follows. First, we begin by exploring some factors conducive to GELLMAI technology Then, we use this context to discuss how these factors are shaping the global race, particularly between the US and China, to develop GELLMAI technology and theorize about implications of this race for cross-society inequalities. Next, we examine how the increasing use of GELLMAI could alter occupational structures and increase income inequality within countries that adopt the technology. Finally, we conclude by situating GELLMAI in the larger historical context of economic production, contrasting the ways in which inequality has been generated in economic activities and transmitted intergenerationally in the past and what may happen in the future.

**GELLMAI Growth Factors**

To understand the social impact of GELLMAI, it is useful to first explore the factors promoting the development of this technology. This discussion will help us anticipate which countries might dominate GELLMAI production–the main factor in determining which nations will bear what economic and sociopolitical consequences–and the rate at which different populations could experience changes in their occupational structure.

*The Scaling Factor*

The first thing to recognize is that GELLMAI is a technology, not a scientific discovery. Technology is characterized by two prominent features: cumulativeness and communality. Firstly, technology is cumulative so that any new technological improvement adds to the extensive reservoir of existing technology. Except for rare instances of secrecy or loss of knowledge, the accumulation of technological inventions grows with time. Secondly, technology is communal in that a new invention benefits not just the inventor, but an entire community. While certain technological advances are sometimes protected as intellectual properties by families, companies, or nations, it is the "best" technology within a community that matters, not the average individual's own technology. Technology is thus communal and inherently shared. Here, the "community" can be defined as a nation-state, a subnational region, or a cluster of nations that share the same language, culture, or political system. In the context of GELLMAI, the size of this community, which we term the scaling factor, is crucial: larger is better. We propose four reasons for this.

Firstly, the size of the community is a relevant factor to the development of GELLMAI technology. In the past, technological invention often stemmed from hard work and trial and error rather than being scientifically derived (Bell 1973). While it is unlikely that any given invention was the result of a purely random event, it is reasonable to assume that, all else being equal, a larger group of people engaged in technological exchange would result in more trials and errors, thereby increasing the likelihood of producing a significant invention for the community.[1] For example, while ancient China did not have science in the same sense it is

---

[1] This aligns with Jared Diamond's (1999) geographical explanation for the development of advanced agricultural economies in Eurasia. Unlike the continents of Africa and the Americas, which are oriented north to south, Eurasia's east-west orientation meant a vast human population sharing similar climates. Since agricultural economies heavily depended on climate, this similarity in climates across Eurasia meant a larger population involved in experimenting with and continually improving agricultural technology through trial and error. At the same time, a

understood in contemporary terms, it excelled in technology, largely owing to its vast population which facilitated numerous trials and errors. Sharing of information within this population was further enabled by China's longstanding written culture.[2] Today, technological advancements are based on modern science rather than simply trial-and-error. Thus, it takes a sufficiently trained labor force to develop GELLMAI technology, but relative to a smaller population with similar levels of education, a larger population can more easily afford a critical mass of scientifically trained workers to meet the demand.

      Second, the larger the community, the more cost-effective it is to develop GELLMAI technology. This principle is a derivative of the classic economic concept known as economies of scale, which suggests that larger production levels allow for lower costs per unit. Developing GELLMAI technology requires significant investment in both the latest computer hardware and in the software implementation of sophisticated data-processing algorithms. A private firm can surmount these cost barriers only in a sufficiently large market. This is especially pertinent due to the "nonrivalry good" (Romer 1990) nature of GELLMAI: in essence, the consumption of GELLMAI technology by additional consumers does not diminish its availability or value for others. Once the technology is developed, the incremental cost for adding additional users is

---

substantially larger population in Eurasia also benefitted from any advancements in agricultural technology in the same land mass.

[2] Since the Qin dynasty (221–206 BCE), administering such a large continent-size country necessitated a unified Chinese writing system for written communication. Local dialects varied so greatly that verbal communication between people from different regions was often impractical. Chinese culture has thus relied on writing.

extremely low, nearly zero. Consequently, firms operating in very large markets can afford the high initial costs associated with developing GELLMAI technology, because it can later recuperate the huge cost from a very large number of consumers. Due to the near-zero marginal cost of consumption and the availability of internet as a mechanism for delivering the technology, a larger community facilitates the consumption of GELLMAI technology.

Third, due to the first key feature of technology – cumulativeness – GELLMAI technology should exhibit a pattern commonly observed in both scientific and technological fields: cumulative advantages. As we have explained, firms serving very large markets are likely to initiate GELLMAI technology development, as they are well positioned to absorb its high costs. However, even after the technology matures and becomes replicable by other firms, those who pioneered it retain an intrinsic advantage – a cumulative advantage.[3] This cumulative advantage arises for two reasons. The first is that knowledge and skills users develop for one GELLMAI implementation are not perfectly transferable to new GELLMAI implementation. That is, once an individual or firm invests time into becoming familiar with a given GELLMAI firm's product, there is a bigger cost to transitioning to new ones. Second, the interactions users have with GELLMAI interfaces are themselves crucial data to improve the technology. Thus, pioneer firms are able to further differentiate their products from those of competitors by taking advantage of the user data. Overall, the initial development of GELLMAI technology, favoring large communalities, leads to a scenario where once successful, these communalities continue to thrive in a self-reinforcing fashion.

---

[3] This phenomenon was described by Robert Merton as "The Matthew Effect," named after a passage in the Book of Matthew (25:29) in the Bible: "For to every one who has will more be given, and he will have abundance; but from him who has not, even what he has will be taken away." In essence, the Matthew Effect dictates that initial success tends to breed further success in a self-reinforcing manner.

Fourth, large and literate communities are proficient in generating substantial amounts of data. Human history hitherto has witnessed three major technological revolutions: the agricultural (ca. 10,000 BCE), industrial (ca. 18th century CE), and information technology (ca. late 20th century CE) revolutions. We are about to experience a fourth technological revolution, the AI revolution. Whereas agriculture depended on land and climate, industry on capital, information technology on human capital, AI relies on vast quantities of data for training and fine-tuning (while still relying, to some extent, on human capital). A society that is both large in population and adequately prosperous can afford both human resources and data.

In summary, this section has established the crucial role of the scaling factor in the development of GELLMAI technology. Economic inefficiency, pragmatic challenges, and a lack of sufficient data are significant hurdles for small societies in developing this technology. Interestingly, the importance of the scaling factor, once critical in agricultural technology but diminished during the industrial era, has regained prominence in the current AI revolution marked by the advent of GELLMAI technology.

*Corpus Specificity and Language Specificity*

GELLMAI systems can generate useful human-like text responses because they are trained on a corpus -- a very large collection of texts -- as input. Therefore, any GELLMAI implementation is necessarily corpus specific. In other words, the technology is as good as the corpus that enables it. This reliance on a specific corpus inherently limits GELLMAI's capabilities. For example, a GELLMAI's accuracy in recounting historical events is confined to the extent and accuracy of training data about the historical events. This means that historical events not well-documented due to reasons like neglect, controversial evidence, or political censorship will not be accurately represented in the model's responses. Moreover, different corpuses can lead to different outputs. This is particularly important when considering the cultural and political context of the corpus. In diverse or international contexts, different corpuses could reflect varying narratives and biases, leading to different responses.

Gender and racial biases in existing English-based GELLMAI technology have been documented (Joyce ed al. 2021; Sun et al. 2023), but differences across languages can be much greater (Luo, Puett, and Smith 2023). To some extent, GELLMAI technology is able to mitigate these issues by varying answers depending on contexts. For example, the same question posed in English might yield a different response when asked in Chinese, reflecting the distinct narratives and contexts inherent in each language. To understand the role of language used, we experimented with ChatGPT 4.0 of OpenAI in December 2023. We asked ChatGPT 4.0 a series of identical questions in four languages: English, Chinese, Japanese, and Burmese. Besides varying languages, we also change the user ethnic identity such as a Chinese or Japanese. Some of the questions we asked are political and cultural, one of which is about a prominent Chinese political leader, and another is about dragon. The experiment yielded the following findings.

1. For concepts and facts that are agreed on across national boundaries, such as scientific terms and discoveries, there is no difference across languages used.

2. For concepts that vary by culture, such as table manners, language matters less than user's identity.[4]

3. For concepts that convey distinctive meanings within languages, such as dragon, input language matters, regardless of self-identification.[5]

4. For terms and concepts that carry different meanings depending on political system or country, language figures prominently. Users get very different answers when politically sensitive terms or concepts are input into ChatGPT in Chinese instead of English. This is surprising because we used the same GELLMAI system -- ChatGPT 4.0.

5. The dissimilarity between answers to questions asked in English and those asked in minor languages such as Burmese are relatively small (although some answers are not even coherent or comprehensible). We surmise that ChatGPT responses in minor languages drew on English-language corpuses.

This last three points arise as a derivative of GELLMAI's corpus specificity, i.e., language specificity, because GELLMAI requires training data -- corpuses – which exist only in specific languages. While GELLMAI technology is, in theory, capable of translating user input into different languages, it performs best in the original language as training data, such as English, because many expressions are unique to particular languages and thus cannot be easily translated into other languages. In other words, translation technology is inherently subject to performance limitations. Consequently, even with identical algorithmic implementations, a GELLMAI model's responses can vary depending on the corpus language as input.

---

[4] Table manners differ significantly across cultures. For example, leaving food unfinished is considered rude in Japan but polite in China, and there is no protocol regarding the matter in the West. Such cultural differences are manifested by ChatGPT 4.0 in different languages as long as the interlocutor's identity is specified.

[5] In Western tradition, dragon has been associated with monstrous mythical creatures. In Chinese culture, dragon (long) has symbolized benevolent power, historically associated with the emperor.

Because corpuses fed into GELLMAI technology are text data in specific languages, languages have an impact on the final GELLMAI products, through what we called earlier "the scale factor." The larger the scale, the more important a language is. We note that, however, a language is not necessarily limited to a single country, such as English, which is spoken in many countries and places that were former colonies of the Great Britain. Conversely, multiple language may be spoken in a country, such as English and French are both official languages in Canada, and there are many official languages in India.

As such, an important factor in the production of GELLMAI technology is the size of a population speaking a given language. Such population sizes vary greatly. In Figure 1, we list the most commonly used languages in the world, with the top being English (at 1.3 billion), followed by Chinese (at 1.1 billion). Although India is currently most populous country in the world, Hindi ranks the third in the language usage.

[Figure 1 About Here]

Like many other social and natural phenomena in the world (Newman 2005), the distribution of language use is highly skewed, following the power law distribution. Few languages, such as English and Chinese, are spoken by a large fraction of people in the world, but most languages are spoken by very few people. In Figure 2, we present a graph illustrating the fit of the population size of language use to the power law which exhibits a Pareto coefficient $\tilde{\alpha} = 1.008$.[6]

[Figure 2 About Here]

As an additional complication, the size of the population speaking a given language does not perfectly predict text data available in that language. For example, while Hindi is the third largest spoken language, a significant portion of the Hindi-speaking population remains illiterate and thus cannot produce text (Statista 2024). In addition, because a large fraction of the educated elites in India are educated in English and communicate in English, Hindi text information does not match its ranking in spoken languages. In terms of newspaper and magazine publications, for example, Hindi ranks fourth; in terms of books, it does not belong to the top 12 (Lobachev 2008). As such, corpus and language specificity will tend to create advantages for linguistic communalities with a large and sufficiently educated population.

**Between-Country Inequality**

As alluded to in the discussion above, it is not clear that we should think about advantages or disadvantages of developing GELLMAI at the country level. After all, the scaling factor and the specificity of corpus creates advantages and disadvantages for linguistic and sociocultural *communities,* which may or may not align with country boundaries. However, for our discussion of the race in GELLMAI technology, it is still useful to treat nations as meaningful units of analysis.

---

[6] Generally, the application of a power law distribution to a given phenomenon is follow by a log-log plot. The log–log plot is asymptotically a straight line with negative slope: $\ln(PX>x)=\ln C-\ln-\ln x$.

The investment in and rise of GELLMAI technology are attributable to its perceived potential to increase economic productivity. As GELLMAI technology becomes more fully developed in the future, we expect meaningful changes in the distribution of the technology. Current GELLMAI business-to-business models take the form of subscription-based enterprise software–that is, firms adopting GELLMAI into their workplace are paying monthly or yearly fees to GELLMAI providers such as OpenAI. As the technology improves and firms adjust their strategies around it, they might reduce employment levels and increasingly automate tasks. Effectively, this can be understood as a form of outsourcing whereby a firm uses a cheaper third-party alternative to perform some of its tasks, increasing its own profits but also that of the service provider. When work is outsourced to a different country, the money that would have been kept within the nation is lost. This is particularly relevant in the case of GELLMAI because companies leading the development of these tools are concentrated in only a few countries and are thus likely to capture a significant share of the money generated by the technology.

In addition to economic concerns, there are sociocultural factors that could fuel between-country inequality. Namely, the need for large corpuses systematically disadvantages people speaking minor languages, as they may be subject to the cultural and political domination of the countries developing GELLMAI systems. The content produced by GELLMAI tools is based on the data it trained on and will thus reflect the attitudes and ideas encoded in those texts and images. For example, around 60% of the training data used for OpenAI's GPT-3 came from the Common Crawl (Brown et al., 2020), a web archive with petabytes of data crawled from the internet. An estimated 46% of documents in the Common Crawl data have English as their primary language and were likely laden with the values of the English-speaking generators of the data (Common Crawl, 2024).

In the decades following the World Ward II, the main theme in world politics has been the celebration of national independence and self-determination (Jackson 2000), free from colonization. The arrival of GELLMAI as a result of the AI revolution now presents a risk to reverse this secular trend, as it may force small countries to be dependent again on dominant ones. In other words, the advent of the AI revolution is likely to increase between-country inequality, favoring large countries with advanced AI technology and disadvantaging small countries lacking independent AI technology. The US-China geopolitical tension and conflict, in particular, could lead to technological competition in the world, making other countries technologically dependent on them.

GELLMAI technology also presents challenges to the current legal systems globally. It has long been accepted that each state has the sovereignty to issue laws within its territory (i.e., Laski 1929). However, as we discussed before, GELLMAI technology will necessarily transcend national boundaries. National differences in such legal domains as data privacy, political censorship, and cross-border data flows will have to be resolved for GELLMAI to be shared cross-nationally. At the current time, Europe could be considered a global leader in data regulation: measures such as the European Commission's *A European Strategy for Data* and the General Data Protection Regulation work together to establish a unified, regulated data market within the EU with the dual goals of ensuring Europe's global competitiveness and data sovereignty (European Commission, 2024). By contrast, the US lacks a federal framework for data regulation, but several states have enacted comprehensive data regulation such as California's Privacy Rights Act and Connecticut's Personal Data Privacy and Online Monitoring

Act. Notably, while only five states had strong data privacy regulations by the end of 2023, 14 more states have signed privacy legislation into law and these are all set to be effective by the beginning of 2026 (Folks, 2024). Meanwhile, the People's Republic of China (PRC) is becoming a significant force in global data regulation. Over the last five years, it has enacted several laws, such as the Cybersecurity Law (CSL), Personal Information Protection Law (PIPL), Data Security Law (DSL), and Measures for Cross-border Data Transfer Security Assessment. These laws aim to build a centrally controlled data governance system, restricting cross-border data flows for national security and public interest reasons and reflecting an increasingly restrictive approach to data regulation (Arroyo et al., 2023). At the same time that the Chinese government restricts cross-border data flows, they have helped stimulate innovation within the country by contracting with Chinese AI firms to process valuable government surveillance data, helping improve their algorithms. For example, Beraja et al., (2022) argue this collaboration between the public and private sector "may have contributed to Chinese firms' emergence as leading innovators in facial recognition AI technology".

**The Case of the US and China**

If the rise of GELLMAI is likely to exacerbate between-country inequality, one important question is which countries are likely to be leaders of these tools and thus enjoy advantages over others. Several scholars and industry leaders have argued that the United States and China are poised to dominate the space, leveraging their extensive resources and strategic investments in AI research and development (see, for example, Graham et al. 2021; Lee, 2018). We examine these two countries in relation to the GELLMAI growth factors outlined above.

For reasons discussed earlier, the US and China benefit from having large populations that use English and Chinese languages. Additionally, vast amounts of written works are published in these two languages. For example, of the 918,964 book titles published worldwide in 1995, the largest number of titles was in English–200,698 titles, at 21.84% of the total– followed by Chinese–100,951 titles, at 10.99% of the total (Lobachev, 2008). Closely related to these figures, China and the US dominated book production. In 2015, China published 470,000 books, and the US published nearly 339,000, with the United Kingdom lagging at a distant third with 173,000 books (International Publishers Association, 2015). Thus, the US and China enjoy advantages in accessing very large corpora available in English and Chinese for training GELLMAI systems.

In terms of technological prowess, the US is a leader in GELLMAI innovation and an originator of the technology. While the exact origin of artificial intelligence is contested, it is clear that US universities played a key role in its creation. Some trace GELLMAI technology back to Alan Turing–a mathematician trained in the University of Cambridge and Princeton University– whose 1950 paper, *Computing Machinery and Intelligence,* explores the mathematical possibility of artificial intelligence and establishes a framework for how to build and test these machines. A few years later, the University of Dartmouth organized the Dartmouth Summer Research Project on Artificial Intelligence, a historic conference where top researchers tested some of Turing's ideas and discussed their visions of the field (Anyoha 2017). The development of neural networks, which are crucial to the statistical training of GELLMAI models, was also borne out of research conducted in US universities, sometimes with funding from government organizations such as the Defense Advanced Research Projects Agency (Anyoha 2017; McKinsey & Company 2018).

In recent decades, state-of-the-art research in GELLMAI technology is also carried out by US-based companies. Indeed, the chess computer, Deep Blue, which famously won a chess match against world champion Garry Kasparov in 1996, was initially developed at Carnegie Mellon University but completed at IBM Research. Years later, Google DeepMind–a British-American research lab–used their innovations in neural network models to win a game of Go against a professional player. Google has also developed products that make significant advances to the field of molecular biology and has published over a thousand papers on GELLMAI research. It is also credited with the creation of transformers, a deep learning architecture that is used in the majority of LLMs. Finally, OpenAI–supported by a large donation from Microsoft in 2019–rose as a prominent leader in the space. Their products include several language models, most famously GPT-3 and GPT-4 which power the popular chatbot and virtual assistant, ChatGPT, which launched in November 2022 and reached 100 million users by January of the following year. OpenAI helped catalyze an "AI Boom," characterized by an exponential growth in investments towards specialized AI companies such as OpenAI and Anthropic and tech giants with a large footprint in the AI space such as Meta, Apple, Alphabet, Amazon, and Microsoft.

Industry leaders in the GELLMAI space are predominantly concentrated in the US, but China is emerging as a formidable competitor to the US[7] (Chou 2023; Kallenborn 2019; Lee 2018; Li, Tong, and Xiao 2021). The AI industry in China is rapidly developing and includes prominent companies such as Alibaba, Baidu, and Tencent. More generally, the sustained and rapid economic development of China since the initiation of its economic reforms in 1978 has significantly boosted its advances in science and technology (Xie, Zhang, and Lai 2014). Notably, by 2020, China had overtaken the US in terms of the volume of scientific papers produced (White 2021). An important aspect of China's science and technology advances has been artificial intelligence, which the government has incorporated into its national agenda since 2006 (He, 2017) and made an explicit goal to become a global AI leader in its thirteenth and fourteenth five-year plans (Luong & Fedasiuk, 2022). Given these developments, it is possible that China could close the gap with the US in GELLMAI technology. Below, we delve into several key factors that are working in China's favor in the realm of GELLMAI technology.

---

[7] Outside of the US and China, there are several other noteworthy companies in the GELLMAI space including Mistral in France, which also develops LLMs. A significant hardware player is Taiwan-based TSMC, which produces an estimated 90% of computer chips required to build these technologies.

(1) <u>A Large Population and Human Capital Base</u>. Currently, China's population is approximately 1.4 billion, over four times the size of the US population. While the per-capita income in China ($13,000 as of 2021) is considerably lower than that in the US ($70,000 as of 2021), the production of college-educated workers in China has been increasing rapidly. In 2019, the number of such individuals was more than double that of the US, with 4 million in China compared to 2 million in the US, as illustrated in Figure 3. A higher proportion of these degrees in China are in science and engineering fields compared to the US (Xie, Zhang, and Lai 2014). Specifically, according to one observer, "China… has a large pool of skilled workers. About 1.4 million engineers qualify annually, six times as many as in the United States, at least a third of them in AI" (Chou 2023).

[Figure 3 About Here]

(2) <u>Favorable Market Conditions for AI Workers</u>. The Chinese labor market highly rewards workers with technical training who are working in technical fields. This contrasts with the labor market in the US, where the highest earnings typically go to practitioners in professions such as law, medicine, and business (Xie, Zhang, and Lai, 2014).

(3) <u>Strong Government Role in Promoting and Investing in AI</u>. The Chinese government prioritizes the development of science and technology as a national strategy to compete internationally and invests heavily in emerging technologies, such as AI (Xie, Zhang, and Lai, 2014). As described above, China has viewed AI as a national priority for over a decade and incorporated it into its five year plans (Kallenborn, 2019). Concentrated and targeted funding to both academic institutions and private businesses developing AI will likely accelerate the development of GELLMAI technology there. According to one estimation, China accounted for 48 percent of global AI venture capital in 2017, surpassing the 38 percent in the United States (Kallenborn, 2019).

(4) <u>A Large Number of Computer Scientists and Engineers of Chinese Descent in the AI Field</u>. Based on last name identification, it has been found that among publications in the top one hundred AI journals and conferences, the proportion of authors with Chinese names nearly doubled from 23.2 percent in 2006 to 42.8 percent in 2015 (Lee and Shehan 2018). A significant portion of these authors are based in the US, either as students or visiting scholars from China, immigrants from China, or as native-born Chinese Americans. They are likely to collaborate with counterparts in China due to cultural affinity (Wang et al. 2012). According to a study by Li, Tong, and Xiao (2021), China has become the dominant player in publishing research papers in the field of AI, accounting for 27.68% of publications in 2017 (37,343 papers), surpassing any other country in the world.

Of course, China also faces challenges in developing GELLMAI technology. We highlight three prominent ones. First, Chinese researchers in the AI field have been making incremental contributions rather than disruptive innovations (Li, Tong, and Xiao 2021). Second, China is now facing US-led restrictions on importing the most advanced computer chips needed for AI deployments (Chou 2023). Third, China lags behind the US in accessing the necessary amount of text data (Lobachev 2008). Can Chinese firms simply use English sources for their GELLMAI implementation? This is unlikely, given China's commitment to a different political system and ideology. On many political and social topics, Chinese perspectives differ, or there is a preference to remain silent. The use of English sources could lead GELLMAI vendors to be at odds with the ideologically laden government. Indeed, in July 2023, the Cyberspace Administration of China, which regulates and censors internet content in the country, announced measures requiring companies to bear legal responsibility for the output of their generative language models and require that this content aligns with Chinese Communist Party ideology (Zhang 2024). In October of the same year, a standards committee published restrictive guidelines on the training data AI companies can use for their models, requiring companies to check that the content does not "incite subversion of state power or of the overthrowing of the socialist system" (The Economist, 2023).

At the same time, the Chinese government recognizes the potential for AI to maintain social stability, allowing the sector to develop rapidly (Beraja et al. 2023). Consequently, the government has adopted more lenient policies regarding data sharing with AI developers to accelerate technological advancement (Beraja et al., 2022; Larsen 2022). In contrast, concerns about legality and individual liberties in the US and Europe have prevented the governments from granting AI companies the same level of data access as seen in China (Larsen 2022).

Regarding their relationships with other technologically advanced countries, the US and China are positioned quite differently in the AI technology sector. Despite holding a large number of important patents, the US does not manufacture many of essential components, such as advanced computing chips for AI technology, and instead relies on international suppliers like

Taiwan-based TSMC (Allison et al. 2021; Chou 2023). While the reliance is underpinned by an international network that has been, to date, stable, the country has enacted policies such as the CHIPS and Science Act in 2022, aiming to incentivize domestic chip manufacturing and strengthen the American semiconductor supply chain. Thus, the US might slowly reduce its dependence on other countries to manufacture essential GELLMAI components (National Science Foundation, 2023). In contrast, China, facing potential technological blockades from the US and other advanced nations, strives to develop all necessary components domestically (Chou 2023; Larsen 2022; Lee and Sheehan 2018). However, in the specific area of computer chip manufacturing, China has yet to succeed in producing the high-speed chips required for advanced AI (Chou 2023) .

In Table 1, we summarize the advantages and disadvantages of the US versus China in dominating the production of GELLMAI technology:

[Table 1 About Here]

**The Impact on the Occupational Structure**

Having discussed the potential for GELLMAI to generate between-country inequality and why China and the US could emerge as leaders in the space, we now turn to an exploration of how this technology could increase inequality within countries, through disruptions in the labor market and changes in the occupational structure.

*Three Lines of Defense for High-Status Jobs*

One pattern we see across multiple professional fields is that, as workers advance through career, their roles change in three ways: (1) tasks performed by people in more senior roles tend to be more varied and involving softer skills (e.g., leadership, communication), and (2) where applicable, these roles have greater exposure to the public or are more likely to involve personal contact with clients, and (3) the work they produce becomes associated with a personal, unique identity. As an example, someone may enter the early stages of an academic career as a research assistant. While research assistants can participate in research projects, their tasks are often well defined, as such cleaning and analyzing data using statistical software, reading and summarizing academic work for literature reviews, etc. For these tasks, one research assistant may be a good substitute for another one. Individuals who progress through an academic career might eventually become professors, directors of research centers, or administrators within a university. These senior roles might include some of the same responsibilities that the research assistant is tasked with (e.g., reading academic work), but these represent only a small share of the full set of tasks these individuals perform in a given week. Thus, advancing in an academic career usually increases the variety of tasks these professionals do on a daily basis and increases the amount of exposure to students or scientists within their academic community and potentially the general public. Additionally, the work of senior academics is deeply intertwined with their unique name and professional identity, giving it a sacrosanct quality. For instance, research labs are frequently named after the principal investigator, symbolizing the connection between the scholar's identity and their work.

      This can be summarized by a simple substitutability principle for AI replacement of job: if a worker can be easily substituted by another worker, it has a high likelihood of being replaced by AI. We highlight these three features of senior roles – task variation, human interaction, and personal identifiability–because all three reduce worker substitutability. As described in Acemoglu and Restropo's (2018) task-based framework for automation, GELLMAI automates

tasks previously done by humans but not necessarily jobs. For this reason, many workers might use GELLMAI in their workplace to handle or speed up some of their tasks without necessarily being replaced by this technology. In our earlier example, a research assistant can be easily substituted, but a professor or a lab director is not. Similarly, senior attorneys at law firms continue having the day-to-day tasks of practicing law (many of which might become automated) but are also tasked with bringing in new clients and are involved with all aspects of the business.

GELLMAI is set to continue improving at exponential rates–indeed, the rate at which AI is improving in key tasks relative to human performance is accelerating (Hutson 2022). As such, these models will rapidly improve in performance in existing tasks and expand the set of tasks it can do, becoming increasingly attractive to employers. Crucially, task-based automation means that workers in jobs that involve a diverse range of tasks are less likely to be displaced in the short term because it is more likely that some of their tasks will remain non-automated. This is especially true if some of their job responsibilities involve communicating with and managing other people ("soft skills" work) as employers might be slower to automate these tasks.

Still, with continuing improvements to GELLMAI, the majority of the tasks currently performed by workers in senior roles such as professors or senior attorneys could eventually be automated. Would this lead to major layoffs for workers in these roles? It is impossible to say for certain, but it seems that the exposure to clients serves as a second insulating factor preventing automation. These jobs are part of a broader category of roles that we expect will maintain a human premium–that is, people will be willing to pay to have the job be completed by a human. In the case of an attorney, even if many of the tasks related to handling a case can be automated, those who can afford it might still prefer face-to-face communication with a human lawyer.

Indeed, recent studies show consumers prefer human customer service representatives to chatbots (Press 2019), human doctors to medical artificial intelligence (Yun et al. 2021), and human counselors to machine services (Ma 2022). The last few decades of tax accounting technology and trends provide evidence of the human premium in action: although ~40 million Americans use technology such as TurboTax to file taxes (Elliot & Kiel 2019), a report by the IRS shows that American taxpayers in income brackets above $75,000 are more likely to use a paid tax return preparer (Internal Revenue Service 2020).

To summarize, workers in senior roles have two-fold protection: first because of the diversity of tasks they are expected to perform and second because their experience and expertise makes them competitive in the eyes of customers willing to pay a human-premium.

*The Threat of Deprofessionalization*

An important note is that, as soon as a task or service is effectively automated, the use of GELLMAI will represent the cheaper option–in the case of TurboTax, for example, the service is free for roughly 37% of taxpayers (TurboTax, 2023). For a consumer to pay the human premium, the worker hired should provide a service that is, in some way, superior to what GELLMAI can provide. In a world where many legal services might be automated, a consumer seeking legal representation may still seek a human attorney, but unlikely for an entry level associate with little experience, who may not represent a large enough improvement to the GELLMAI service to warrant the additional cost. Selective senior lawyers who have already built a reputation for their expertise remain highly competitive and can charge a human premium, but those that are only starting out or have not achieved the same level of success face a higher threat.

As a result, workers performing lower-level roles in these professions are vulnerable because they do not enjoy the protections afforded to high-status workers. Instead of disappearing professions, we may observe disappearing *ladders*–jobs that used to mark the culmination of a progression through an occupational ladder of increasingly higher-level occupations. While high-level positions are protected from full automation, the traditional rungs (paralegal, research assistant, etc.) are not. The disappearing of ladders is not in itself a new development–for years, social scientists have documented the shrinking share of the American adult population that is middle income (see Kochhar et al. (2015) for a review) and the polarizing of the US labor market into high-wage and low-wage jobs (Autor et al. 2015). However, the rise of GELLMAI could accelerate this trend, threatening jobs in middle and upper middle-income strata and pushing Americans further into the two extremes of the income distribution.

In the absence of traditional occupational ladders and assuming senior level roles remain in demand, how will these positions be filled once current senior workers exit due to retirement? One possibility is *deprofessionalization* – a reduction of large number of especially trained and skilled workers to perform specific tasks -- a phenomenon we have seen in other occupations throughout history. A classic example is painting, which was once regarded as skilled labor and had a well-established occupational ladder where aspiring painters underwent rigorous apprenticeships under master artists. However, as societal attitudes and needs towards art changed, along with the development of photographic and later video-recording technologies, the occupational ladder largely disappeared. Aspiring painters today have to build their portfolios through a combination of formal education, independent practice, and personal project, investing their own resources, both time and money, to create a body of work that showcases their abilities. This example provides some evidence of deprofessionalization*;* although professional painters still exist, those aspiring to obtain that role must train and create projects without compensation, often treating it as a hobby outside of their main source of income. While it might seem improbable at this moment that jobs in accounting, medicine, or law could take a similar path, just as the invention of photography deprofessionalized portrait painters, new technology such as GELLMAI could lead to such drastic occupational restructuring.

*Implications for Higher Education*

Resource inequality among institutions of higher education is known to have been rising (Xie 2014). The penetration of GELLMAI in colleges and universities will likely further intensify institutional inequality to new high levels. This is because today's teaching functions of colleges and universities can be performed by GELLMAI in the future, as some knowledge can be stored, retrieved, and taught by GELLMAI. Most instructors of higher education may not be able to compete with GELLMAI for effectiveness and thoroughness in teaching known knowledge. In addition, the arrival of the AI revolution may dramatically reduce the demand for college-level knowledge, because knowledge per se will be no longer as valuable as today in the labor market, a topic to be discussed in the next section. At the same time, college affordability has become a key issue in the US (Baker, 2024) and institutions seeking to reduce costs could turn towards GELLMAI technology to provide student instruction. Indeed, over the last few decades, universities have dramatically increased the use of part-time, adjunct instructors to reduce expenses (Bettinger & Long, 2010), and the use of GELLMAI would represent a next step in their effort to increase efficiency. Thus, we might see a deprofessionalization of college teachers in the near future.

However, the value of elite and highly prestigious universities will remain high or even increase in the new AI-dominant era. We conjecture this to be true for three reasons. First, with the widespread use of AI technology to manufacture goods and provide services, new technology creators and new knowledge producers, who are likely to be employed by elite universities and are highly differentiated from each other, are highly valued. Second, elite universities will likely shift their instructional emphasis from imparting known knowledge to knowledge creation, creative use of knowledge, and the improvement of non-cognitive (soft) skills. Third, with education of knowledge being secondary, human relationship and social identity will become increasingly important. Universities can fulfill the needs for human bonding among students and give them a strong sense of social identity and belonging, but universities with primarily online instruction or without a robust student community residing near campus may be less equipped to provide this. In the end, we may see an enhanced role for elite universities and fierce competition for first-rate researchers, but the need for traditional classroom instruction could decline and reduce demand for ordinary teachers.

*The Potential Role of Organized Labor*

The displacement of middle-income jobs by AI can be resisted by organized labor. A historic example of this was the 2023 Writers Guild of America (WGA) strike, one of the longest labor disputes in the history of US entertainment. Union leaders representing more than 11,000 screenwriters sought to increase payment and job security for writers and explicitly demanded limits to the use of AI for entertainment content production. The eventual agreement between the WGA and the Alliance of Motion Picture and Television Producers contractually prevents the use of LLMs to write scripts or to use GELLMAI output as source material and prohibits use of writers' material as training data, effectively reducing the possibility for screenwriting to become automated in the near future.

The WGA strike began just months after OpenAI released ChatGPT and represented one of the first large-scale labor conflict between human workers and GELLMAI. However, organized labor and collective-bargaining agreements have long played a role in protecting workers from displacement due to technological advances. One early example comes from the railroad industry which underwent a rapid transition from steam to diesel locomotives in the decades following World War II. The new diesel-based technology completely restructured the railroad work force and eliminated the need for firemen who tended the fire to power a steam engine. However, the Brotherhood of Locomotive Firemen and Engine was a strong union that protected firemen by demanding firemen to be included in diesel crews (Klein, 1990). While the union eventually lost a long legal battle to railroad companies that rendered the fireman's job as no longer necessary, they were able to protect these jobs for 26 years after railroads fully transitioned to diesel, allowing many firemen to seek training to become railroad engineers or employment elsewhere (Chicago Tribune, 1985). Similarly, the development of technologies such as letter sorting machines, optical character readers, and barcode sorters affected clerk and mail handlers in the postal service, but the American Postal Workers Union fought for a "no layoff" contract clause that prevented these workers from losing their job (Rubio, 2020).

While these examples illustrate the power of collective bargaining, successfully preventing or alleviating job displacement has historically depended on strong labor unions and strong federal labor regulations upholding workers' rights. Yet, union membership peaked in 1950 and steadily declined in the decades that followed (Feiveson, 2023), leaving many workers today with relatively less bargaining power to counter the effects of GELLMAI-induced automation. In addition, a more globalized economy has made it easier for companies to outsource labor to other countries, further weakening worker power. Nevertheless, the share of Americans approving of unions in 2022 and 2023 hovered around 70%, higher than it has been since the 1960s (Saad, 2023). The labor movement also saw historic wins in 2023, with workers participating in the largest number of strikes in recent history (AFL-CIO, 2024). It is thus possible that we will see a resurgence of the labor movement that could prevent or at least slow down job losses.

**The Big Picture**

Confucius once said, "Study the past if you would define the future." While we cannot observe the future, we can draw upon the past to gain educated glimpses about the future. In this section, we review the major historical stages in terms of primary technology for economy and draw a big picture for the possible scenario in the future when and if the AI fundamentally transforms the world. Continuing our earlier discussion, besides the impact on economy, we also discuss technological implications for social inequality and social mobility. We summarize our main argument in Table 2, beginning with the primitive, hunter and gatherer economy.

[Table 2 About Here]

As far as we know, social inequalities have always existed in every human society, including the most primitive ones. However, the primary basis for social inequality has evolved alongside technological advancements. In a hunting and gathering economy, social inequality existed but in a less pronounced than later economies (Smith et al. 2010; Smith, Smith, and Codding 2021). Wealth and status were based on individual differences in food yield, which largely depends on "considerable strength and stamina, visual acuity, and other aspects of good health" (Smith et al. 2010, p. 21). Production was for immediate subsistence, as there was no storage of food, and mobility was high, preventing people from staying in one place for long. As a result, there was little accumulation of wealth or possessions (Smith et al. 2010, p. 21) and the transmission of social advantages and disadvantages was thus limited, primarily occurring through genetics and luck.

Following the agricultural revolution, the agricultural economy became characterized by permanent settlement, human organization, and the rise of inequality (Davis, 2018). With the advent of agriculture, humans began to own private property, the most significant of which was land. As agricultural production is heavily dependent on land, land ownership became the primary basis for social inequality as of land intergenerational transmission of social advantages and disadvantages took the form inheritance. Later, the industrial revolution brought along machines that replaced human and animal power as the main sources of production (Bell 1973; Stearns 2020). In the industrial economy, manufactured goods become abundant, improving the standard of living beyond subsistence for the first time in history (Clark 2007). For a relatively small number of capitalists, the ownership of capital became a source of income, known as property income (Piketty 2014). For most people, however, the operation of machines formed the basis of labor income. Intergenerational transmission of social advantages and disadvantages in this economy thus took the form of skill transfers and capital inheritance.

Most recently, we have been experiencing a post-industrial economy known as the knowledge economy. This concept is extensively discussed by Bell in his landmark 1973 book, *The Coming of Post-Industrial Society*. The primary output of the knowledge economy is services. Much of the routine work is replaced by computers, and knowledge becomes increasingly important. This is evident as many professional services, such as legal, health, financial, and educational sectors, require specialized knowledge. Consequently, parents are strongly incentivized to invest in their children's education as a form of intergenerational transfer. Of course, for the small minority who are wealthy, capital remains an important factor.

Now, as AI technology continues to grow, we may arrive at a post-knowledge economy that will bring with it new forms of intergenerational transmissions of inequality. We conjecture that GELLMAI will render the possession of knowledge less important in the labor market. Not only the manufacturing of goods but also the provision of services will be automated by AI-powered machines so most people may not need to work many hours, as machines can perform tasks on their behalf. If the production of these technologies continues to be concentrated in a few countries, this could increase the economic dependence of smaller countries on nations like the US and China and potentially lead to pernicious forms of cultural domination. In addition to between-country inequality, societies within countries could also become deeply divided, with a tiny minority occupying elite positions and working long hours while the vast majority contribute little directly to the production of goods and provision of services. This could bring with it a problem of disappearing ladders, where traditional occupational career paths are disrupted, and the labor market becomes even more deeply polarized.

In the future AI economy, what might matter the most? For a small minority, capital and ownership of AI technology as means of production remain important and can be passed on to the next generation. For workers already in high-status positions, the diversity of tasks they are expected to perform, their high exposure to clients, and the fact that their work may be uniquely associated with their identity serve as job protections. Workers in low-income positions who perform manual, or person-to-person service jobs also face a less immediate threat from AI and might not experience the shocks of this economic transition as acutely. However, many workers in middle income jobs are already feeling the impact of these new technologies and are at higher risk of replacement. For these workers, personality and soft skills might begin to play outsized roles for their labor market outcomes. Workers exposed to GELLMAI are likely to be valued for how effectively they use AI and how they present themselves to others. Individuals will be uniquely valued for who they are to others. Social, as well as personal, identity will be paramount. So will be personal ties. To "sell" oneself effectively, soft skills will become extremely important. Thus, much of the intergenerational transmission of social status may take the form of these soft skills.

**Conclusion**

GELLMAI will most likely grow in importance and fundamentally transform humanity in ways we cannot fully anticipate at this point. Given the potential for these new tools to exacerbate the already growing levels of inequality in countries like the US and China, it is of utmost importance to create policies that regulate these technologies and counteract their possibly harmful distributional effects. In 2024, the US Department of Labor announced a new set of principles meant to provide guidance for employers seeking to adopt GELLMAI technology to "enhance job quality and protect workers' rights" (Department of Labor, 2024). Although such guidance is a useful step, designing effective policies to guide the GELLMAI transition is hard to do at the federal level given that each industry—and indeed, each firm—will have very particular automation needs, making their actions hard to regulate. Organized labor could also play a crucial role in minimizing job losses and shielding workers from some of the more harmful effects of automation, but the labor movement faces steep challenges such as right-to-work laws and anti-union tactics from employers that make organizing difficult. To ensure a smoother transition into the AI economy, countries would be well-advised not just to regulate GELLMAI technology but also to strengthen their laws protecting unions to ensure a healthy balance of power.

There is a scale factor to the development of GELLMAI technology, according large countries advantages relative to others. The US and China are current leaders in the GELLMAI space and will continue to leverage their edges over smaller countries. Given the corpus specificity and associated language specificity of GELLMAI technology, the two countries will provide services to other countries with content reflecting different political systems and cultures. We anticipate fierce competition between the US and China for dominance of GELLMAI technology, as the stakes are large—indeed, global.

We speculate about the forthcoming arrival of a post-knowledge society as a result of the AI Revolution. If products and services can be readily provided by AI-powered machines, we anticipate large-scale displacement of jobs. Displacement is particularly likely to occur for workers who are now considered the middle class, such as teachers, accountants, clerks, computer programmers, engineers, editors, doctors, and lawyers. Workers both at the very top and at the bottom of the social hierarchy are less likely to be displaced. Knowledge and hard skills will become less important, but soft skills will increase in importance. In this future AI-driven society, people will care less about the material conditions (such as quality) of products and services, as they will be little differentiated due to AI, but more about *who* provides products and services. In other words, personal identity will gain significance. In shopping for products and services, people will be swayed less by objective criteria than by personal subjective tastes. Individuals and companies will be successful not for meeting other people's material needs but their psychological needs—making them happy and satisfied.

As with the other technological advances described earlier, GELLMAI has the potential to boost economies and increase standards of living by reducing the cost of goods and allowing workers more time to pursue personal interests, engage in creative endeavors, and contribute to their communities. However, as we argue in this paper, this technology is also poised to increase between- and within-country inequality if the transition to an AI-driven society is not managed carefully. Proper government regulation is crucial to ensure ethical standards, mitigate risks, and foster an inclusive environment where the benefits of AI are widely shared.

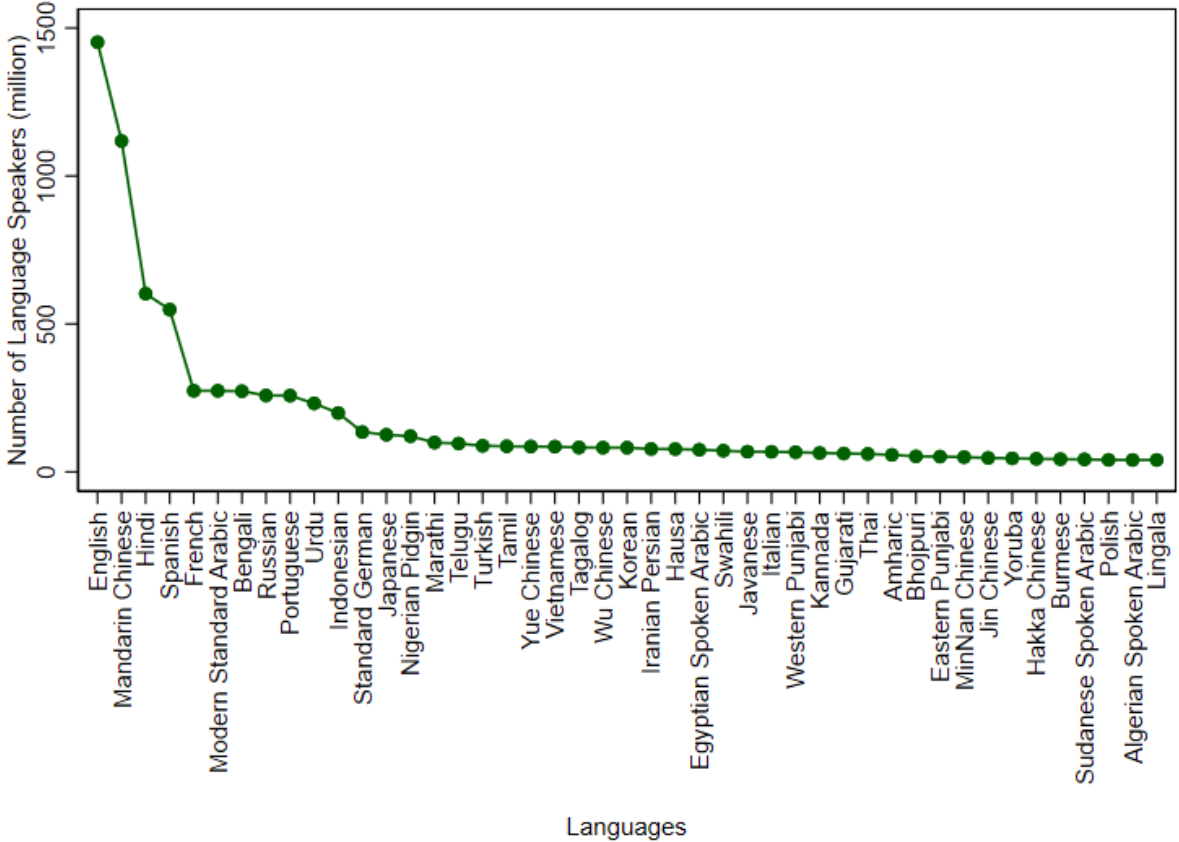

Figure 1: Population Size by Language.

Data source:

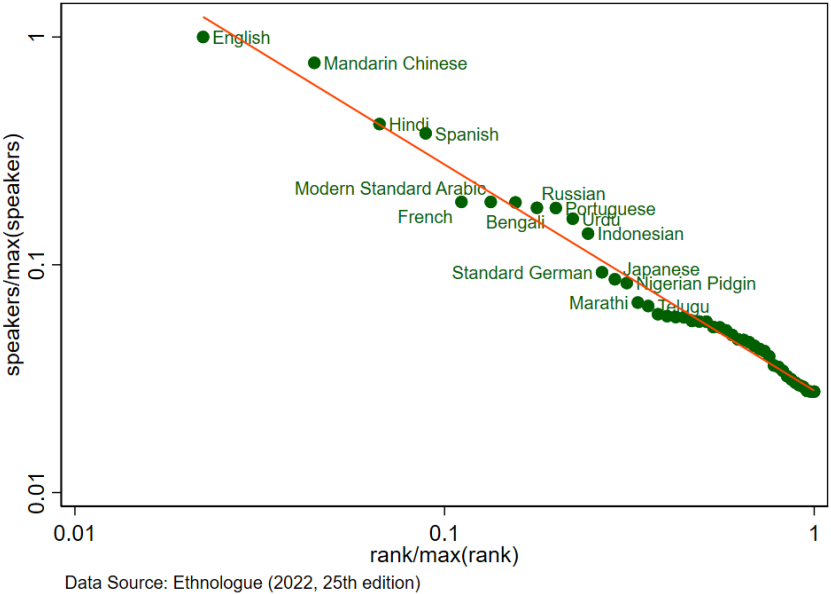

Figure 2: Power-Law Distribution of Population Size by Language

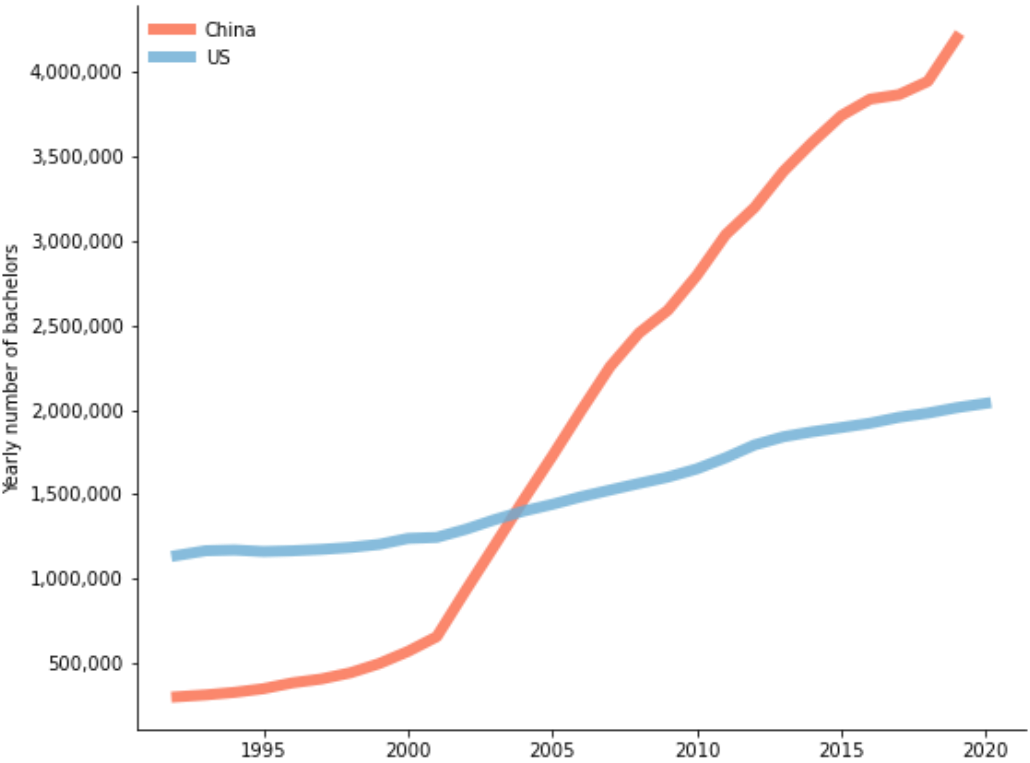

Figure 3: Trends in production of bachelor's degrees 1991-2019, the US and China Compared.

Data source:

Table 1: Comparison of the US and China for the Dominance of AI

| Country | Relative Advantages | Relative Disadvantages |
| --- | --- | --- |
| The US | More disruptive innovation | Smaller AI labor force |
| | Best hardware | More legal restrictions to user data |
| | Large text data base | Little government support |
| | International network for collaboration | Lack of domestic supply-chain |
| China | Larger AI labor force | No disruptive innovation |
| | Access to larger user data bases | More restricted text data |
| | Strong government support | Limited hardware access |
| | Strong domestic supply chain | Lack of international allies |

Table 2: Economy Type and Associated Characteristics

| Economy Type | Alternative Name | Primary Production | Basis for Social Inequality | Means of Intergenerational Transmission |
|---|---|---|---|---|
| Hunting and gathering | Pre-agricultural | Subsistence | Physical strength and skills | Genetic |
| Agricultural | Agricultural | Food | Land | Land |
| Industrial | Post-agricultural | Manufactured goods | Operation of machines | Capital and manual skills |
| Knowledge | Post-industrial | Services | Knowledge possession | Capital and education |
| Personalized | Post-knowledge | Personal interaction | Personality | Capital and Soft skills |